\documentclass[%
 aip,
 amsmath,amssymb,
preprint,
]{revtex4-1}

\usepackage{graphicx}
\usepackage{dcolumn}
\usepackage{bm}
\usepackage[utf8]{inputenc}
\usepackage[T1]{fontenc}
\usepackage{mathptmx}

\begin{document}

\preprint{}

\title[]{Stress distribution at the AlN/SiC heterointerface probed by Raman spectroscopy}

\author{I.D. Breev}
  \email{breev.ilia.d@mail.ioffe.ru}
\author{K.V. Likhachev}%
\author{V.V. Yakovleva}
\affiliation{ 
Ioffe Institute, Russia, Saint-Petersburg, Politechnicheskaya st., 26
}%
\author{R. H\"ubner}
\author{G.V. Astakhov}
\affiliation{Institute of Ion Beam Physics and Materials Research,
Helmholtz-Zentrum Dresden-Rossendorf (HZDR), Germany, Dresden, Bautzner Landstr. 400}
\author{P.G. Baranov}
\homepage{http://www.ioffe.ru/labmsc/ru/main.html}
\author{E.N. Mokhov}
\author{A.N. Anisimov}
 \email{aan0100@gmail.com}
\affiliation{ 
Ioffe Institute, Russia, Saint-Petersburg, Politechnicheskaya st., 26
}%

\date{\today}

\begin{abstract}
We grow AlN/4H-SiC and AlN/6H-SiC heterostructures by physical vapor deposition and characterize the heterointerface with nanoscale resolution. Furthermore, we investigate the spatial stress and strain distribution in these heterostructures using confocal Raman spectroscopy. We measure the spectral shifts of various vibrational Raman modes across the heterointerface and along the entire depth of the 4H- and 6H-SiC layers. Using the earlier experimental prediction for the phonon-deformation potential constants, we determine the stress tensor components in SiC as a function of the distance from the AlN/SiC heterointerface. In spite that the lattice parameter of SiC is smaller than that of AlN, the SiC layers are compressively strained at the heterointerface. This counterintuitive behavior is explained by different coefficients of thermal expansion of SiC and AlN when the heterostructures are cooled from growth to room temperature. The compressive stress values are maximum at the heterointerface, approaching one GPa, and relaxes to the equilibrium value on the scale of several tens of microns from the heterointerface.
\end{abstract}

\maketitle

\section{\label{sec:level1}Introduction}
In recent years, the investigation of wide-bandgap semiconductor heterostructures (AlN, SiC) with different refractive indices  has become of great interest for power electronic devices \cite{hitemp} and photonic crystals in the ultraviolet spectral range \cite{N_el_2014}. For such objectives, it is highly important to exactly understand the properties of the material interfaces grown using various techniques. The Raman scattering method is of great use for non-destructive analysis of such structures for the determination of stress values and overall crystal quality \cite{PhysRevB.58.12899,ProcHetero}. 
Furthermore, SiC hosts spin qubits \cite{Son:2020kh}, particularly silicon vacancies \cite{Kraus:2013di} and divacancies  \cite{Falk:2013jq}, which can be coherently controlled at room temperature \cite{Koehl:2011fv, NatureComm}, possess a long coherence time in the ms range  \cite{Christle:2014ti, Simin:2017iw}, reveal single-photon emission  \cite{Christle:2014ti, Widmann:2014ve, Fuchs:2015ii} with a spectrally narrow zero-phonon line  \cite{Christle:2017tq, Morioka:2020iv}, and show integrability into electronic and photonic circuits  \cite{Fuchs:2013dz, Anderson:2019kh, Lukin:2019fh, Niethammer:2019ir}. The parameters of these spin qubits, including the zero-field splitting and zero-phonon line are inhomogeneously broadened due to local variations of charge and strain in SiC crystals \cite{Baranov:2011ib, Riedel:2012jq, Simin:2016cp, Nagy:2019fw, Banks:2019je, Nagy:2019fw}. Because AlN is a strong piezoelectric material, it may allow to control the local strain at the AlN/SiC heterointerface through applying voltage. Importantly, the refractive index of AlN $(n = 2.1)$ is less than that of SiC $(n = 2.6)$. This can potentially allow the fabrication of SiC-on-AlN photonic cavities with light confinement in SiC and therefore enhanced interaction of light with spin qubits.
 
To realize the aforementioned concepts, a high-quality AlN/SiC heterointerface is of crucial importance. In the previous studies, the Raman spectroscopy of AlN layers grown on SiC substrates has been performed to optimize the growth protocols \cite{semicond}. Thick AlN/SiC heterostructures were investigated along the entire depth of the AlN layers \cite{ConfSer}. However, the properties of SiC close to the  AlN/SiC heterointerface have not been investigated so far. 

We study the Raman peak positions associated with the  $E_{2}$ TO (x=1/2), $A_{1}$ TO (x=1), $E_{1}$ TO (x=0) phonon modes in 4H-SiC and $E_{2}$ (x=1/3), $E_{2}$ TO (x=1), $E_{1}$ TO (x=0), $A_{1}$ TO (x=0) phonon modes in 6H-SiC \cite{Nakashima}. The SiC Raman peaks at the heterointerface demonstrate clear spectral shifts towards higher energies, pointing at compressing stress of the lattice. This result is counterintuitive, because the in-plane lattice parameter of SiC is smaller than that of AlN and one could expect that the SiC lattice tends to laterally expand, resulting in tensile stress. The opposite behavior is related to the different coefficients of thermal expansion of AlN and SiC, as explained in section~\ref{Discussion}.

Using phonon-deformational potentials in 4H-SiC \cite{Sugie}, we determine the stress tensor components across the AlN/4H-SiC interface and along the entire depth of the 4H-SiC layer. In addition, we estimate the stress value in 6H-SiC across the AlN/6H-SiC interface and along the entire depth of the 6H-SiC layer using the pressure dependence of the Raman peaks \cite{SiCRaman}.

\section{Experimental setup and sample growth\label{Expsetup}}
To perform spatially resolved Raman scattering experiments, we used a linearly polarized laser with a wavelength  $\lambda$=532nm and power P=5mW. The laser was focused using a 100x objective with a numerical aperture NA=0.9. In the confocal arrangement with a pinhole of 100 $\mu$m, it  provides an in-plane resolution (along the \textbf{Z} and \textbf{X} directions) of 1$\mu$m.  A precise \textbf{XYZ} piezoelectric scanner and \textbf{XZ} in-plane micromechanical driver were used to position the sample. The Raman spectra were recorded with a thermoelectrically cooled CCD camera and a monochromator equipped with a 2400-grooves-per-mm  grating and a linear polarizer. 

\begin{figure}
\includegraphics{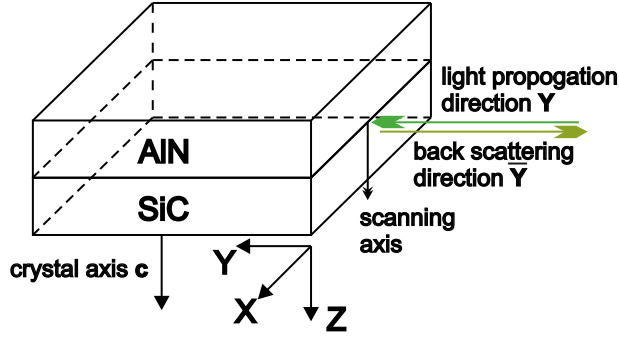}
\caption{\label{geometry} Experimental geometry scheme showing the sample orientation with crystal axis \textbf{c}, Cartesian coordinate system axes \textbf{XYZ}, light propagation direction \textbf{Y} and back scattering direction $\overline{\textbf{Y}}$.}
\end{figure}

To define the experimental geometry, we introduced a cartesian coordinate system \textbf{XYZ}. The sample is investigated in the back-scattering geometry with the light propagation direction \textbf{Y} and the back-scattering direction $\overline{\textbf{Y}}$. The crystal axis \textbf{c} is parallel to \textbf{Z}, and the laser polarization direction \textbf{E} can be either parallel to \textbf{X} or \textbf{Z}. The polarization direction axis \textbf{P} can be oriented either parallel to \textbf{X} or \textbf{Z}. The samples were not specifically oriented in the \textbf{XY} plane. Thereby, the available Raman scattering geometries for our experiments were $\textbf{Y(X;X)}$$\overline{\textbf{Y}}$, $\textbf{Y(X;Z)}$$\overline{\textbf{Y}}$, $\textbf{Y(Z;Z)}$$\overline{\textbf{Y}}$, and $\textbf{Y(Z;X)}$$\overline{\textbf{Y}}$. The corresponding experiment geometry showing the heterostructures orientation with the crystal axis \textbf{c}, the \textbf{XYZ} cartesian axis orientation, the light propagation direction \textbf{Y}, and the back-scattering direction $\overline{\textbf{Y}}$ is depicted in Fig.~\ref{geometry}. All Raman spectra were collected with 20 seconds accumulation time.

To visualize the AlN/6H-SiC heterointerface, high-angle annular dark-field scanning transmission electron microscopy (HAADF-STEM) imaging and spectrum imaging based on the energy-dispersive X-ray spectroscopy (EDXS) were performed at 200 kV with a Talos F200X microscope equipped with an X-FEG electron source and a Super-X EDX detector system (FEI). Prior to STEM analysis, the specimen mounted in a high-visibility low-background holder was placed for 2 s into a Model 1020 Plasma Cleaner (Fischione) to remove possible contamination. Cross-sectional TEM specimen preparation of the AlN/6H-SiC interface region was done by in situ lift-out using a Zeiss Crossbeam NVision 40 system. To protect the area of interest, a carbon cap layer was deposited beginning with electron-beam-assisted and subsequently followed by Ga focused-ion-beam- (FIB) assisted precursor decomposition. Afterwards, the TEM lamella was prepared using a 30 keV Ga FIB with adapted currents. Its transfer to a 3-post copper lift-out grid (Omniprobe) was done with a Kleindiek micromanipulator. To minimize sidewall damage, Ga ions with only 5 keV energy were used for final thinning of the TEM lamella to electron transparency.

The AlN crystal was grown on 4H- and 6H-SiC substrates by physical vapor deposition, specifically the sublimation sandwich method (SSM)\cite{CrystalGrowth2,CrystalGrowth3}. The main growth temperature was 2000 $\, ^\circ$C, the distance between source and substrate was in the range of 3-10 mm, the temperature gradient was 5 K/mm, the nitrogen pressure in the growth chamber was 0.3-1 atm. For a more detailed description of the SSM method, the reader is referred to Ref.~\onlinecite{CrystalGrowth}.
The AlN/4H-SiC heterostructure is characterized by AlN and 4H-SiC layer thicknesses of 228/242 $\mu$m, respectively, while the AlN/6H-SiC heterostructure has AlN and 6H-SiC layer thicknesses of 253/420 $\mu$m, respectively.
\section{Experimental results}
\subsection{\label{AlN/4H-SiC sample Raman spectroscopy}Raman spectroscopy of the AlN/4H-SiC heterostructure}
\begin{figure*}
\centering
\includegraphics{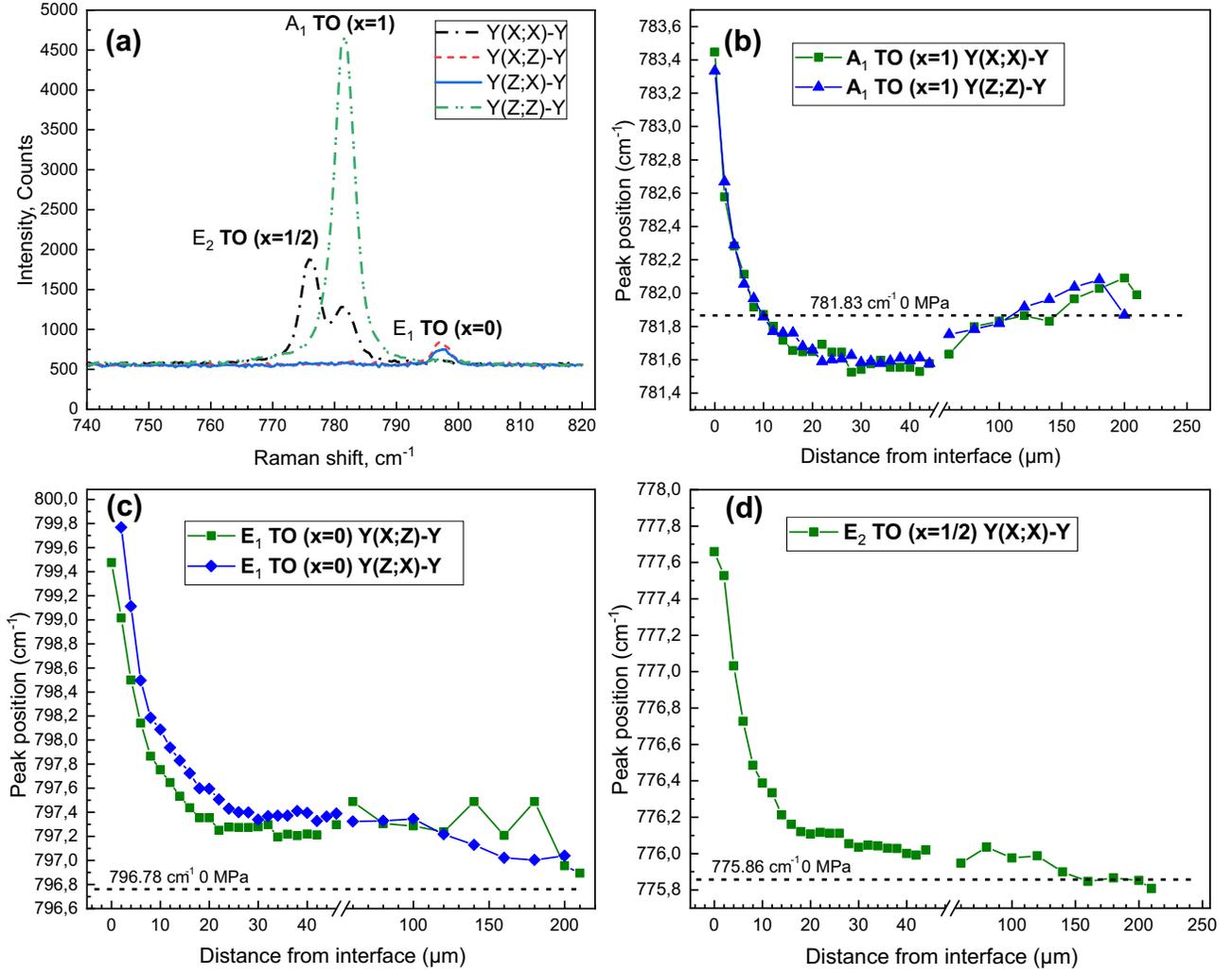}
\caption{\label{4H-SiCAnalisys}(a) Raman spectra of 4H-SiC included in the AlN/4H-SiC heterostructure for 4 Raman scattering geometries depicted in different colors in the TO Raman region. The corresponding phonon modes for each Raman peak are shown. (b, c, d) 4H-SiC Raman peaks behavior across the AlN/4H-SiC interface and along the entire depth of the 4H-SiC layer for (b) $A_{1}$ TO (x=1), (c) $E_{1}$ TO (x=0), (d) $E_{2}$ TO (x=1/2) phonon modes.}
\end{figure*}

We studied the Raman peaks evolution in 4H-SiC across the interface and along the entire depth of the 4H-SiC layer for 4 different Raman scattering geometries. Our main interest is in the TO Raman region (760-810 $cm^{-1}$).
In Fig.~\ref{4H-SiCAnalisys} (a), Raman spectra of the 4H-SiC layer included in the 4H-SiC/AlN heterostructure are shown for 4 Raman scattering geometries: $\textbf{Y(X;X)}\overline{\textbf{Y}}$ (black dashed one-dot line), $\textbf{Y(X;Z)}\overline{\textbf{Y}}$ (red dashed line), $\textbf{Y(Z;Z)}\overline{\textbf{Y}}$ (green dashed two-dots line) and $\textbf{Y(Z;X)}\overline{\textbf{Y}}$ (blue solid line). There are 3 Raman phonon modes in the TO Raman region: $E_{2}$ TO (x=1/2), $A_{1}$ TO (x=1), and $E_{1}$ TO (x=0), where $E_{2}$, $A_{1}$, and $E_{1}$ are symmetries of phonon modes in Mulliken notation, TO is the transversal optic phonon type, and x is the reduced wave vector of the phonon modes in the basic Brillouin zone\cite{Nakashima}. The corresponding phonon modes are shown for each Raman peak. 
We also obtained 4H-SiC Raman spectra across the AlN/4H-SiC interface and along the entire depth of the 4H-SiC layer. Then, we performed a fitting procedure of the 4H-SiC Raman peaks in the TO region by Voigt functions. In Fig.~\ref{4H-SiCAnalisys} (b, c, d), the Raman peaks positions for the AlN/4H-SiC heterostructure are presented for the 4 Raman scattering geometries. We performed the analysis for the phonon modes $A_{1}$ TO (x=1) (Fig.~\ref{4H-SiCAnalisys}(b)), $E_{1}$ TO (x=0) (Fig.~\ref{4H-SiCAnalisys}(c)), and $E_{2}$ TO (x=1/2) (Fig.~\ref{4H-SiCAnalisys}(d)). The unstressed values of the Raman peak positions were measured for the standard 4H-SiC sample (Table~\ref{4H-SiCtheory}) and are shown in Fig.~\ref{4H-SiCAnalisys} (b, c, d) by dashed black lines.
We observe that Raman peak behavior of 4H-SiC in Fig.~\ref{4H-SiCAnalisys}(b,c) obtained in different Raman geometries coincides well. However, the small experimental discrepancy ~0.2 $cm^{-1}$ between the Raman peak positions for the $E_{1}$ TO (x=0) phonon mode results in a significant error of about 100 MPa in the stress tensor component $\sigma_{in-plane}$ which was determined by the calculations to be discussed in chapter \ref{Theory}.

\subsection{\label{3.2.  AlN/6H-SiC interface Raman spectroscopy.}Raman spectroscopy of the AlN/6H-SiC heterostructure}

\begin{figure*}
\centering
\includegraphics{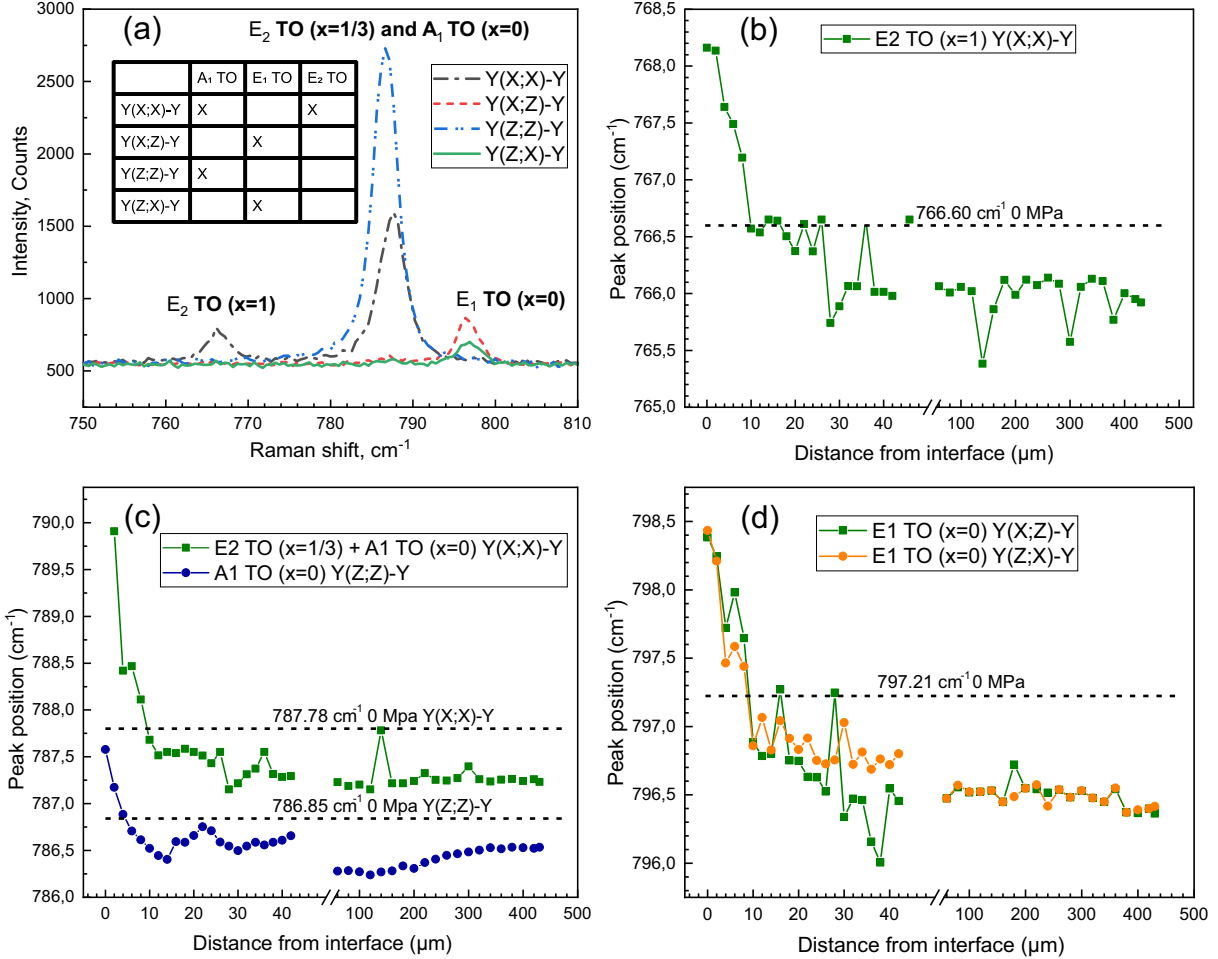}
\caption{\label{6H-SiCAnalisys}(a) Raman spectra of 6H-SiC included in the AlN/6H-SiC heterostructure for 4 Raman scattering geometries. The activity of the phonon modes in the different Raman geometries is shown in the table inside the graph. (b, c, d) 6H-SiC Raman peak behavior across the AlN/6H-SiC interface and along the entire depth of the 6H-SiC layer for (b) $E_{2}$ TO (x=1) modes (c) double-degenerated 788 $cm^{-1}$ Raman peak consisting of the phonon modes $E_{2}$ TO (x=1/3) and $A_{1}$ TO (x=0), and (d) $E_{1}$ TO (x=0).}
\end{figure*}

Similarly, we investigated the Raman peaks evolution in 6H-SiC across the AlN/6H-SiC interface and along the entire depth of the 6H-SiC layer for 4 different Raman scattering geometries. In Fig.~\ref{6H-SiCAnalisys} (a), Raman spectra of the 6H-SiC layer included in the AlN/6H-SiC heterostructure are shown for the 4 Raman scattering geometries: $\textbf{Y(X;X)}\overline{\textbf{Y}}$ (black dashed one-dot line), $\textbf{Y(X;Z)}\overline{\textbf{Y}}$ (red dashed line), $\textbf{Y(Z;Z)}\overline{\textbf{Y}}$ (blue dashed two-dots line), and $\textbf{Y(Z;X)}\overline{\textbf{Y}}$ (green solid line). The activity of the phonon modes in different Raman geometries is shown in the table inside the graph, which is determined by the Raman polarization rules for the crystallographic symmetry $C^{4}_{6v}$ of 6H-SiC \cite{bilbao}. We observe 4 Raman phonon modes in the TO Raman region: $E_{2}$ TO (x=1), $E_{2}$ TO (x=1/3), $A_{1}$ TO (x=0), and $E_{1}$ TO (x=0), whereby the notations are the same as in part~\ref{AlN/4H-SiC sample Raman spectroscopy}. We face the problem that the 788 $cm^{-1}$ Raman peak is double-degenerated to the phonon modes $E_{2}$ TO (x=1/3) and $A_{1}$ TO (x=0). Following Refs.~\onlinecite{SiCRaman},\onlinecite{oldRaman} and the Raman polarization rules for the $C^{4}_{6v}$ symmetry\cite{bilbao}, we conclude that in the $\textbf{Y(X;X)}\overline{\textbf{Y}}$ geometry, the 788 $cm^{-1}$ Raman peak is double-degenerated, and in the $\textbf{Y(Z;Z)}\overline{\textbf{Y}}$ geometry, it includes only the $A_{1}$ TO (x=0) mode. As for 4H-SiC, the unstressed values of the Raman peaks positions were measured for a standard 6H-SiC sample and are shown in Fig.~\ref{6H-SiCAnalisys} (b, c, d) by dashed black lines.

Likewise, we obtained 6H-SiC Raman spectra across the AlN/6H-SiC interface and along the entire depth of the 6H-SiC layer. Then, we performed a fitting procedure of the 6H-SiC Raman peaks in TO region by Voigt functions. In Fig.~\ref{6H-SiCAnalisys}, the results of the 6H-SiC Raman peak behavior in the AlN/6H-SiC heterostructure are presented for the 4 Raman scattering geometries (Fig.~\ref{6H-SiCAnalisys}(b,c,d)). In particular, we performed the analysis for the phonon modes $E_{2}$ TO (x=1) (Fig.~\ref{6H-SiCAnalisys}(b)), the double-degenerated 788 $cm^{-1}$ Raman peak consisting of the phonon modes $E_{2}$ TO (x=1/3) and $A_{1}$ TO (x=0) (Fig.~\ref{6H-SiCAnalisys}(c)), and $E_{1}$ TO (x=0) (Fig.~\ref{6H-SiCAnalisys}(d)).


\subsection{\label{XDS}Transmission electron microscopy analysis of the AlN/6H-SiC heterostructure interface}
\begin{figure}
    \centering
    \includegraphics{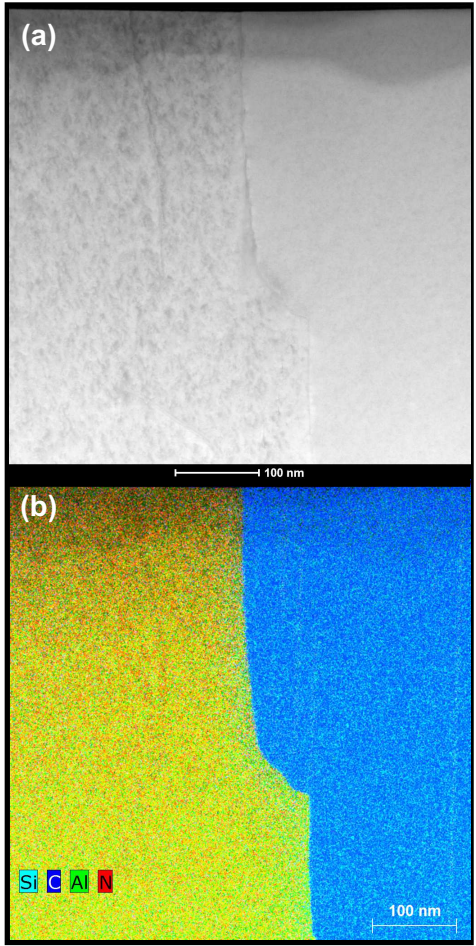}
    \caption{HAADF-STEM image (a) and superimposed element distributions based on EDXS analysis (b) for the AlN/6H-SiC heterostructure.}
    \label{fig:XDS}
\end{figure}

To gain further insights into the properties of the interface between AlN and 6H-SiC, we analyzed the AlN/6H-SiC heterostructure by means of scanning transmission electron microscopy (STEM) coupled with element mapping based on energy-dispersive X-ray spectroscopy (EDXS). While HAADF-STEM imaging allows for the localization of the interface (Fig.~\ref{fig:XDS}(a)), the spatial distribution of the chemical composition on the nanometer scale is provided by EDXS analysis (Fig.~\ref{fig:XDS}(b)). Although the AlN/6H-SiC interface shows a certain morphological roughness, there are no indications for a solid solution of AlN and SiC directly at the interface. 

\section{\label{Theory}Calculations}
\subsection{\label{4H-SiCtheory}4H-SiC stress tensor evaluation}
\begin{table}
\caption{\label{tab:Phonon}Phonon-deformation potentials for 4H-SiC\cite{Sugie}}
\begin{ruledtabular}
\begin{tabular}{llll}
Phonon symmetry & Peak frequency & $a'_i$ & $b'_i$ \\

&$(cm^{-1})$ & $(cm^{-1}/GPa)$ & $(cm^{-1}/GPa)$ \\
\hline
& Our experimental data & &\\
\hline
$E_{2}$ TO (x=1/2) & 775.86 & -1.55 & -0.74\\
$A_{1}$ TO (x=1) & 781.83 & -0.46 & -2.67\\
$E_{1}$ TO (x=0) & 796.78 & -2.06 & -0.43\\
\end{tabular}
\end{ruledtabular}
\end{table}

\begin{figure*}
\centering
\includegraphics{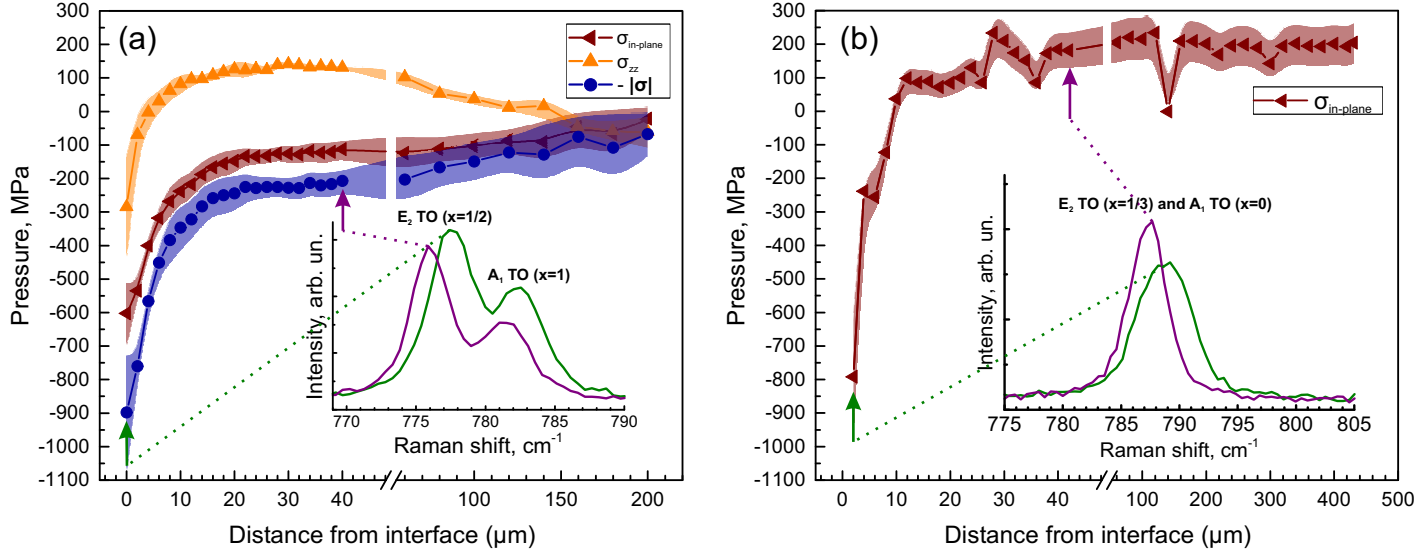}
\caption{(a) The stress tensor components $\sigma_{in-plane}$ (brown triangles), $\sigma_{zz}$ (orange triangles), and the minus stress tensor module \textbf{$-|\sigma|$} (dark blue circles) distribution along the entire depth of the 4H-SiC layer inside the AlN/4H-SiC heterostructure with error bars depicted as background-colored areas. The inset presents the 4H-SiC Raman spectrum shift near the interface and in depth. (b) The stress tensor component $\sigma_{in-plane}$ (brown triangles) distribution along the entire depth of the 6H-SiC layer inside the AlN/6H-SiC heterostructure with error bars depicted as background light-brown area. The inset presents the 6H-SiC Raman spectrum shift near the interface and in depth.}
\label{StressConclusion}
\end{figure*}

After the Raman peak position measurements in the investigated heterostructures, we were able to estimate the stress value and deepen our knowledge of the interface's state and quality. To perform stress distribution calculations, we use theoretical information on the Raman spectra modes and Raman peak positions without stress in 4H- and 6H-SiC\cite{Nakashima,oldRaman}. In our analysis, we neglected the shear stress components $\sigma_{xy}=\sigma_{xz}=\sigma_{yz} = 0$. We also assumed that the stress in the sample $\textbf{c}$-plane is biaxially isotropic $\sigma_{in-plane}= \sigma_{xx}= \sigma_{yy}$. This procedure was taken from Ref.~\onlinecite{Sugie} to preserve the logic of the phonon-deformation potentials obtainment procedure. Consequently, the stress tensor $\sigma$ becomes diagonal, i.e. has only the 3 nonzero components $\sigma_{xx,yy,zz}$. With these assumptions and the phonon-deformation potentials for 4H-SiC (Table~\ref{tab:Phonon}), we conclude that the Raman peak shifts have the following form:
\begin{eqnarray}
    \Delta w_i = w_i - w_{0i}; \nonumber\\
    \Delta w_i = 2a'_i \sigma_{in-plane} + b'_i\sigma_{zz}; \\
    \Delta w_j = 2a'_j \sigma_{in-plane} + b'_j\sigma_{zz};\nonumber
\end{eqnarray}
where \textit{i},\textit{j} are different phonon modes, for example $E_2$ TO (x=1/2), $A_1$ TO (x=1), or $E_1$ TO (x=0), $w_i$ is a Raman peak position of phonon mode \textit{i} under mechanical stress, $w_{0i}$ is a Raman peak position of phonon mode \textit{i} in the absence of stress, $a'_{i,j},b'_{i,j}$ are the phonon-deformation potentials of the phonon modes \textit{i} and \textit{j}. Having 2 equations and 2 unidentified variables, we calculate:
\begin{eqnarray}
\sigma_{in-plane}=\frac{\Delta w_i - \frac{b'_i}{b'_j}\Delta w_j}{2a'_i-2a'_j\frac{b'_i}{b'_j}};\nonumber\\
\sigma_{zz}=\frac{\Delta w_i - \frac{a'_i}{a'_j}\Delta w_j}{b'_i-b'_j\frac{a'_i}{a'_j}};
\end{eqnarray}
The stress tensor components in the 4H-SiC layer were calculated using the Raman shifts of the different phonon modes. It should be noted that for a better estimation of the $\sigma_{in-plane}$ and $\sigma_{zz}$ stress components, we used 3 different pairs of Raman peak shifts: $E_2$ TO (x=1/2) and $A_1$ TO (x=1) modes, $E_1$ TO (x=0) in ($\textbf{Y(X;Z)}\overline{\textbf{Y}}$ geometry and $A_1$ TO (x=1) modes, $E_1$ TO (x=0) in ($\textbf{Y(Z;X)}\overline{\textbf{Y}}$ geometry and $A_1$ TO (x=1) modes. After that, we averaged the calculation results over three points and determined the error. The outcome is presented in Fig.~\ref{StressConclusion}(a) for the $\sigma_{in-plane}$ (brown triangles), $\sigma_{zz}$ (orange triangles) and $-|\sigma|$ (dark blue circles), which was calculated as:
\begin{eqnarray}\label{module}
-|\sigma|=-\sqrt{tr(\sigma \sigma^T)}=-\sqrt{2\sigma_{in-plane}^2+\sigma_{zz}^2}
\end{eqnarray}
Also, the inset in Fig.~\ref{StressConclusion}(a) shows the 4H-SiC Raman spectrum shift near the interface and in depth. The maximum in absolute value in-plane stress component $\sigma_{in-plane}$ in the 4H-SiC layer near the interface is about -600 MPa and drops by the factor of a \textit{e} at a distance of approximately 6 $\mu$m from the interface, which we derived from exponential fitting.

\subsection{\label{6H-SiCtheory}6H-SiC stress evaluation}
There is no complete set of phonon deformation potentials for 6H-SiC in literature. However, using the proportionality coefficient between 6H-SiC 788 $cm^{-1}$ peak position and the stretching pressure on $(11\overline{2}0)$ crystal face experimentally obtained in Ref.~\onlinecite{SiCRaman}, we took the liberty to roughly estimate the stress tensor component $\sigma_{in-plane}$ in the AlN/6H-SiC heterostructure. The researchers in Ref.~\onlinecite{SiCRaman} proposed that the $E_2$ TO (x=1/3) mode Raman peak is partially concealed by the $A_1$ TO (x=0) mode Raman peak, which coincides with data given in Ref.~\onlinecite{oldRaman,bilbao}. We used that proportionality coefficient to estimate the stress value $\sigma_{(11\overline{2}0)}$:
\begin{eqnarray}
\sigma_{(11\overline{2}0)} = -0.52 \Delta w_{comb};\nonumber\\
\Delta w_{comb} = w - w_0;
\end{eqnarray}
where $w_{comb}$ is the Raman peak position under stress, corresponding to the $E_2$ TO (x=1/3) phonon mode together with $A_1$ TO (x=0) phonon mode and $w_0$ = 787,78 $cm^{-1}$ is the Raman peak position without stress, which we experimentally determined for standard 6H-SiC sample. We neglected all sheer stress components $\sigma_{xy}=\sigma_{xz}=\sigma_{yz} = 0$, following the procedure from paragraph~\ref{4H-SiCtheory}. Since we didn't have the phonon-deformation potential for Z axis, we had to neglect $\sigma_{zz}$ = 0 stress tensor component, what is the main factor of our estimation inaccuracy. However, this obstacle did not prevent us from a general trend estimation. Since $\sigma_{(11\overline{2}0)}$ is associated with the total stress, i. e. is equal to $-|\sigma|$, we derived the stress tensor component $\sigma_{in-plane}$ for the comparison with the 4H-SiC stress calculation results, using equation~(\ref{module}) as:
\begin{eqnarray}
\sigma_{in-plane} = \frac{-|\sigma|}{\sqrt{2}} = \frac{\sigma_{(11\overline{2}0)}}{\sqrt{2}}
\end{eqnarray}
The resulting stress tensor component $\sigma_{in-plane}$ estimation is depicted in Fig.~\ref{StressConclusion}(b). The inset shows the 6H-SiC Raman spectrum shift near the interface and in depth. The maximum in absolute value in-plane stress component $\sigma_{in-plane}$ in the 6H-SiC layer near the interface is about -800 MPa and drops by a factor of \textit{e} at a distance of approximately 4 $\mu$m from the interface, which we derived from exponential fitting.

\subsection{\label{4H-SiC strain}4H-SiC stress to strain conversion}
For better usefulness in the fields of spin and optical properties calculations, we decided to convert the obtained stress tensor components into equivalent strain tensor values. Generally, for such purpose, we need the elastic compliance tensor $\textbf{S}$. However, it is easy to use and obtain the elastic stiffness tensor $\textbf{C}$ for SiC, whose components are given in Table~\ref{tab:Ctensor}\cite{Ctensor}. The relationship between the stress tensor $\sigma$ and the strain tensor $\epsilon$ , representing the generalized Hooke's law, can be expressed as:
\begin{eqnarray}
\sigma =\textbf{C}\epsilon;\nonumber\\
\sigma_{xx} = C_{11}\epsilon_{xx}+C_{12}\epsilon_{yy}+C_{13}\epsilon_{zz};\nonumber\\
\sigma_{yy} = C_{12}\epsilon_{xx}+C_{11}\epsilon_{yy}+C_{13}\epsilon_{zz};\\
\sigma_{zz} = C_{13}\epsilon_{xx}+C_{13}\epsilon_{yy}+C_{33}\epsilon_{zz};\nonumber
\end{eqnarray}
To perform stress to strain conversion, we had to take several factors into consideration . As the $\sigma$ stress tensor, in our system, the $\epsilon$ strain tensor has only 3 nonzero components, i.e. $\epsilon_{xx}$, $\epsilon_{yy}$, and $\epsilon_{zz}$. Also, given that $\sigma_{xx}=\sigma_{yy}=\sigma_{in-plane}$, we have $\epsilon_{xx}=\epsilon_{yy}=\epsilon_{in-plane}$, and the relationships transform into:
\begin{eqnarray}
\label{sigmaepsilon}
\sigma_{in-plane} = (C_{11}+C_{12})\epsilon_{in-plane}+C_{13}\epsilon_{zz};\\
\sigma_{zz} = 2C_{13}\epsilon_{in-plane}+C_{33}\epsilon_{zz};\nonumber
\end{eqnarray}
Deriving $\epsilon_{in-plane}$ and $\epsilon_{zz}$ from equations~(\ref{sigmaepsilon}), we obtained the $\epsilon$:
\begin{eqnarray}
\epsilon_{in-plane} = \frac {C_{33}\sigma_{in-plane}-C_{13}\sigma_{zz}}{C_{11}C_{33}+C_{12}C_{33}-2C_{13}^2};\\
\epsilon_{zz} = \frac {2C_{13}\sigma_{in-plane}-(C_{11}+C_{12})\sigma_{zz}}{2C_{13}^2-C_{33}(C_{11}+C_{12})};\nonumber
\end{eqnarray}

Using the $\textbf{C}$ elastic stiffness tensor components from table~\ref{tab:Ctensor} and the stress tensor $\sigma$ components for the 4H-SiC layer of the AlN/4H-SiC heterostructure from Fig.~\ref{StressConclusion} (a), we calculate $\epsilon_{in-plane}$ and $\epsilon_{zz}$ values.

\begin{figure}
    \centering
    \includegraphics{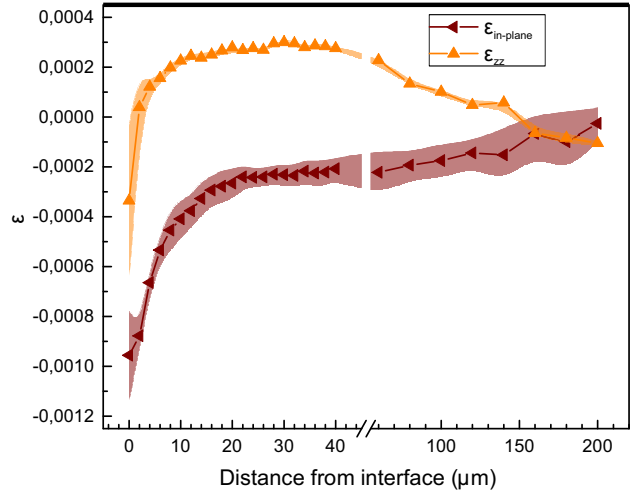}
    \caption{Strain tensor components $\epsilon_{in-plane}=\epsilon_{xx}=\epsilon_{yy}$ (brown triangles), $\epsilon_{zz}$ (orange triangles) evolution with distance from the AlN/4H-SiC interface inside the 4H-SiC layer calculated using stress tensor components.}
    \label{Epsilontensor}
\end{figure}

The results of this calculations are presented in Fig.~\ref{Epsilontensor} for the $\epsilon_{in-plane}$ (brown triangles) and $\epsilon_{zz}$ (orange triangles) strain tensors components together with error bars depicted as background-colored areas.
 
\begin{table}
\caption{\label{tab:Ctensor} The elastic constants in unit of GPa of 6H and 4H SiC at room temperature.\cite{Ctensor}}
\begin{ruledtabular}
\begin{tabular}{llllll}
$\textbf{C}$ tensor component & $C_{11}$&$C_{33}$&$C_{44}$&$C_{12}$&$C_{13}$\\
\hline
GPa&$501\pm 4$&$553\pm 4$&$163\pm 4$&$111\pm 5$&$52\pm 9$\\
\end{tabular}
\end{ruledtabular}
\end{table}
 
\section{\label{Discussion}Discussion}
\par We assume that the main factor contributing to the stress formation at the AlN/SiC heterointerface is the difference in the coefficients of thermal expansion between AlN and SiC. The contribution to the mechanical stress of the silicon carbide surface is excluded. First, in the growth process, the SiC surface is etched away and the damaged layer is removed, where permanent deformation can potentially be formed. Second, we investigated the SiC substrates before the AlN growth and no Raman shifts were found in them. The AlN layers are grown on SiC substrates at 2000 $\, ^\circ$C, and during the growth process, the crystal is strain free.  When the samples are cooled from the growth temperature to room temperature, the AlN crystal shrinks larger than SiC. For both SiC polytypes, the change in the lattice parameter $a$ in the plane of the interface is smaller than that for AlN (see data in Table \ref{LatticePar}). As a result, the SiC crystal is compressed at the heterointerface relative to a stress-free SiC crystal. In turn, it does not allow the AlN crystal to shrink to the equilibrium value (as for the unstressed AlN crystal), and the AlN crystal at the heterointerface is stretched. The latter is difficult to confirm using the Raman spectroscopy only. This is because of the high degree of alloying and inhomogeneity of the AlN crystal~\cite{semicond}. The tensile stress of the same order is observed in the buffer layers of AlN on SiC \cite{Zollner}. The mechanism of thermal compression in AlN/SiC heterostructures by observing the AlN Raman modes far from the AlN/SiC heterointerface has been described \cite{Liu}. Thus, the detailed analysis of the strain distribution in AlN is beyond the scope of this work, and we concentrated on SiC. We found that the stress value and the characteristic relaxation length in SiC varies with the relative thicknesses of the SiC and AlN. It is about 6 $\mu$m for the AlN/4H-SiC heterosctructure and is about 4 $\mu$m for the AlN/6H-SiC heterosctructure.
\par It should also be noted that the deformation of the 4H-SiC lattice reaches a maximum at the heterointerface of approximately $\epsilon_{in-plane}$ = -0.001  (according to  Fig.~\ref{Epsilontensor}), corresponding to the change in the lattice parameter $\Delta a_{hetero}$ = -0.00308  {\AA}. When the heterostructure cooled down, the equilibrium lattice parameter decreases by 0.02617 {\AA} for 4H-SiC and by 0.04155 {\AA} for AlN (see data in Table \ref{LatticePar}). Thus, the AlN unit cell should shrink by $\Delta a_{thermal}$ = -0.01538 {\AA} more than the SiC unit cell in the interface plane upon cooling. Thus, the actual compression of the 4H-SiC crystal lattice at the heterointerface is about $\Delta a_{hetero} / \Delta a_{thermal}$ = 20 \%  with respect to the difference in the expected compressions for AlN and 4H-SiC. We attribute this to the fact that upon cooling not all crystalline bonds between AlN and 4H-SiC are retained \cite{Tanaka}.

\begin{table}
\caption{\label{LatticePar} Crystal lattice parameters \textbf{a} for room and growth temperatures calculated for 6H-SiC\cite{6Hlattice}, 4H-SiC\cite{4Hlattice} and AlN\cite{AlNlattice}}
\begin{ruledtabular}
\begin{tabular}{llll}
T, K & 4H-SiC & 6H-SiC & AlN \\
& a, \AA & a, \AA &  a, \AA \\
\hline
300 & 3.08352 & 3.084732583 & 3.11561223 \\
2100 & 3.109693 & 3.111591236 & 3.15716127 \\
\end{tabular}
\end{ruledtabular}
\end{table}

\section{\label{Conclusion}Conclusion}
We performed Raman spectroscopy on two heterostructures AlN/4H-SiC and AlN/6H-SiC. We also got an insight into the AlN/6H-SiC interface properties with nanoscale resolution using HAADF-STEM and EDXS analysis. We investigated the 4H- and 6H-SiC Raman peak behavior across an AlN/SiC interface and along the entire depth of the SiC layers. We determined the stress tensor component evolution in the 4H-SiC layer and estimated the in-plane stress value for the 6H-SiC layer. The maximum stress tensor component $\sigma_{in-plane}$ value lies in the range from -500 to -700 MPa (Fig.~\ref{StressConclusion} (a)) for the AlN/4H-SiC heterostructure and is about -800 MPa (Fig.~\ref{StressConclusion} (b)) for the AlN/6H-SiC heterostructure. The conducted research can illuminate the quality of the interface of AlN/SiC-wideband semiconductor heterostructures, their internal structure, and behavior. Our estimations can provide necessary information for AlN/SiC power electronic devices and fabrication of ultraviolet photonic crystals. Also, this is a clear demonstration of the Raman scattering application for the stress measurements in SiC of 4H- and 6H- polytypes. Furthermore, we envision a possibility for spin qubit control in SiC by stress and strain using the piezoelectrical properties of the AlN underneath.
\begin{acknowledgments}
The authors thank I.A. Eliseev for consultation on the topic of the Raman polarisation rules in 6H-SiC.
The authors thank A. Kunz for TEM specimen preparation and Y. Berenc{\'e}n for the Raman test measurements before the preparation. Furthermore, the use of the HZDR Ion Beam Center TEM facilities and the funding of TEM Talos by the German Federal Ministry of Education of Research (BMBF), Grant No. 03SF0451, in the framework of HEMCP are acknowledged. This work is supported by the Russian Science Foundation under grant No. 19-72-00154.
\end{acknowledgments}

\section*{Data availability}
The data that support the findings of this study are available from the corresponding author upon reasonable request.

\nocite{*}

\bibliography{aipsamp}

\end{document}